\documentstyle{mn}

\title{The influence of binary stars on the kinematics of low-mass
  galaxies}

\author[S. De Rijcke \& H. Dejonghe]{S. De Rijcke$^{1, 2,*}$ and H. Dejonghe$^2$ \\
  $^1$ Current address~: Astronomisches Institut, Universit\"at Basel,
  Venusstrasse 7, 4102 Binningen, Switzerland \\
  $^2$ Sterrenkundig Observatorium, Ghent University, Krijgslaan 281,
  S9, 9000 Gent, Belgium \\
  $^*$ Postdoctoral Fellow of the Fund for Scientific Research -
    Flanders (Belgium)(F.W.O)\\
  {\tt sven.derijcke@rug.ac.be} \\
  {\tt herwig.dejonghe@rug.ac.be} }
\begin{document}

\maketitle

\begin{abstract} 
  In this paper, the influence of binary stars on the measured
  kinematics of dwarf galaxies is investigated. Using realistic
  distributions of the orbital parameters (semi-major axis,
  eccentricity, \ldots), analytical expressions are derived for the
  changes induced by the presence of binary stars in the measured
  velocity moments of low-mass galaxies (such as the projected
  velocity dispersion and the 4$^{\rm th}$ order Gauss-Hermite
  coefficient $h_4$). It is shown that there is a noticeable change in
  the observed velocity dispersion if the intrinsic velocity
  dispersion of a galaxy is of the same order as the binary velocity
  dispersion. The kurtosis of the line-of-sight velocity distribution
  (LOSVD) is affected even at higher values of the intrinsic velocity
  dispersion.
  
  Moreover, the LOSVD of the binary stars (i.e. the probability of
  finding a star in a binary system with a particular projected
  velocity) is given in closed form, approximating the constituent
  stars of all binaries to revolve on circular orbits around each
  other. With this binary LOSVD, we calculate the observed LOSVD, the
  latter quantifying the movement of stars along the line of sight
  caused both by the stars' orbits through the galaxy and by the
  motion of stars in binary systems. As suggested by the changes
  induced in the moments, the observed LOSVD becomes more peaked
  around zero velocity and develops broader high-velocity wings. These
  results are important in interpreting kinematics derived from
  integrated-light spectra of low-mass galaxies and many of the
  intermediate results are useful for comparison with Monte Carlo
  simulations.

  \end{abstract}

\section{Introduction}

The usual approach to derive the kinematics (mean velocity along the
line of sight, the velocity dispersion, \ldots) of galaxies from
observed spectra is comparing them to the spectra of so-called
template stars. The broadening of the absorption lines in the galaxy
spectra is then interpreted as a result of the orbital motion of the
myriad of stars along the line of sight through the galaxy~: each star
has a Doppler shifted spectrum and the absorption lines consequently
appear at slightly different wavelengths. These template stars are
carefully selected~: they should not rotate too rapidly, not be member
of a binary system or be peculiar in any other way so as to be a good
representation of the average stellar population. Thus, the broadening
of the absorption lines is explained completely in terms of stellar
orbital motions.  However, a large fraction of the stars in a galaxy
are members of a binary (or even multiple) system. Stars in binary
systems orbit the center of mass of the two stars and this extra
velocity adds to the velocity dispersion of the galaxy. Hence, the
measured kinematics of galaxies are not independent of the population
of binary stars.

Previous authors have used Monte Carlo simulations to estimate the
influence of binaries on the observed velocity dispersion (see
Hargreaves {\em et al.} \cite{har} and references therein). This work
focused mainly on observations of dwarf galaxies in the Local Group or
globular clusters that can be resolved into individual stars. For a
sample of stars in a galaxy, radial velocities are measured and from
these velocities, the mean velocity and velocity dispersion profiles
can be established. The Monte Carlo calculations by construction are
very suited to interpret these ``discrete'' data, e.g. it is possible
to check the effect of repeated observations to weed out short-period
binaries. The crowded inner regions of globulars or dwarf galaxies
outside of the Local Group however are impervious to this kind of
study.  There, integrated-light spectra are needed to measure the
kinematics.  As bigger telescopes make it possible to penetrate to
ever lower levels of surface brightness and hence to study low-mass
stellar systems at large distances, it is important to investigate the
effect of binary stars on the derived LOSVD. Since Monte Carlo
calculations at present are unable to yield the full LOSVD or even
higher order moments such as the kurtosis without excessive
computational effort, they cannot be employed in this context. Hence,
we give analytical expressions for the velocity moments of the binary
stars and for the LOSVD itself (approximating all binaries to consist
of stars on circular orbits).
\section{Definitions}

We first give a number of definitions of important quantities and
clarify some notations. We focus our attention on a single binary
system. One star, the ``primary'', has a mass $M$ and the other, the
``secondary'' has a mass $m$. All orbital parameters pertain to the
primary's orbit~:
\begin{itemize}
\item $v_p$ : the velocity of the primary with respect to the binary's
  center of mass, projected onto the observer's line of sight
\item $\mu_p$ : the gravitational constant of force for the primary's
  orbit around the center of mass. If $\mu = G(m+M)$ is the force
  constant of the orbit relative to the secondary, then
\begin{eqnarray}
  \mu_p = \left( \frac{m}{m+M} \right)^3 \mu = G \frac{m^3}{(m+M)^2}.
  \label{defmu}
\end{eqnarray}
\item $a_p$ : semi-major axis of the primary's orbit with respect to
  the binary's center of mass. If $a$ is the semi-major axis of the
  primary's orbit with respect to the secondary, then
\begin{eqnarray}
a_p = \left( \frac{m}{m+M} \right) a. \label{defa}
\end{eqnarray}
\item $T, \,e, \, i, \,\omega, \,\Omega,$ : respectively the period,
  eccentricity, inclination, argument of periastron and longitude of
  ascending node of the primary's orbit. The latter does not affect
  the primary's position projected onto the line of sight nor its
  line-of-sight velocity and hence does not enter the calculations.
\item $\phi$ : the true anomaly along the primary's orbit.
\end{itemize}
\begin{figure}
\vspace{6.5cm}
\special{hscale=35 vscale=35 hsize=570 vsize=250 
         hoffset=-25 voffset=200 angle=-90 psfile="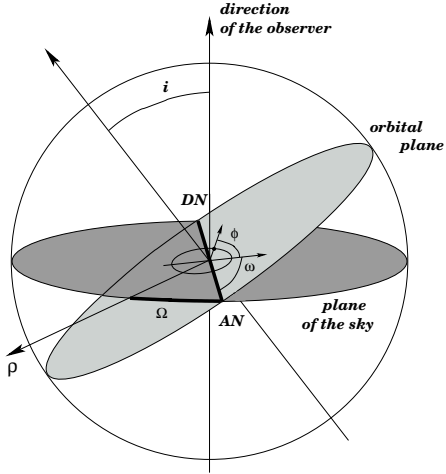"}
\caption{The orbit of the primary around the binary's center of mass. 
  Indicated are the descending node $DN$ and the ascending node $AN$,
  the longitude of ascending node $\Omega$ with respect to some
  reference direction $\rho$, the argument of the periastron $\omega$,
  the true anomaly $\phi$ and the inclination $i$. \label{orbitps}}
\end{figure}
\begin{figure}
\vspace{6.8cm}
\special{hscale=40 vscale=40 hsize=570 vsize=250 
         hoffset=-15 voffset=220 angle=-90 psfile="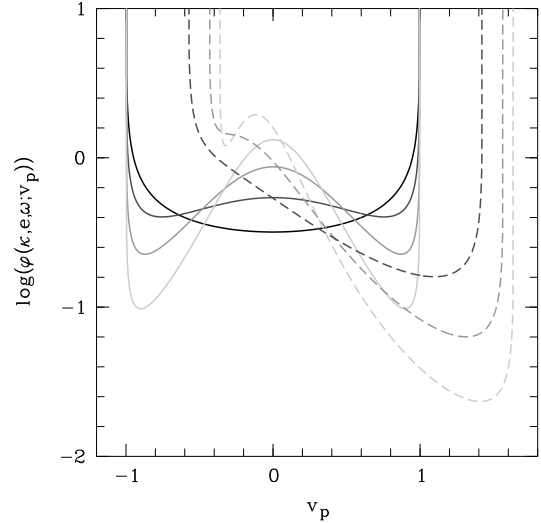"}
\caption{The specific binary LOSVD $\varphi(\kappa,e,\omega;v_p)$. 
  In full lines, the LOSVD is plotted for $\omega=90^\circ$ and $e=0$
  (black), $e=0.6$ (dark grey), $e=0.8$ (grey) and $e=0.9$ (light
  grey). The other parameters are set by the choice $\kappa = 1$. In
  dashed lines, the LOSVD is plotted for $\omega=45^\circ$ and the
  same eccentricities. 
  \label{phi_intr} }
\end{figure}

\section{The specific binary LOSVD}

The specific binary LOSVD gives the probability of finding the primary
star with mass $M$ with a line-of-sight velocity in the interval
$[v_p-\Delta v_p/2,v_p+\Delta v_p/2]$, given that it revolves around a
secondary star with mass $m$ on an elliptical orbit with orbital
parameters $a$, $e$, $\omega$ and $i$. Intuitively, this probability is
proportional to the fraction of its period during which the primary
has a velocity in this interval. Mathematically, this translates into
the following expression for the specific binary LOSVD~:
\begin{equation}
\varphi(\kappa,e,\omega;v_p)\,dv_p = \frac{1}{T} \frac{dv_p}{\left| \frac{dv_p}{dt} \right|}. \label{ansatz}
\end{equation}
with $v_p$ the line-of-sight velocity of the primary at phase angle
$\phi$ on its orbit
\begin{equation}
v_p = \kappa ( e \cos \omega + \cos (\phi + \omega) ). \label{velo}
\end{equation}
Here,
\begin{equation}
\kappa = \frac{2 \pi a_p \sin i}{T \sqrt{1-e^2}} = \sqrt{ \frac{\mu_p}{a_p}} 
\frac{\sin i }{\sqrt{1-e^2}}.
\end{equation}
Hence, if follows that
\begin{equation}
  \left| \frac{dv_p}{dt} \right| = \frac{2 \pi \kappa}{(1-e^2)^{3/2}
    T} \left| \sin(\phi+\omega) \right| ( 1 + e \cos \phi)^2
\end{equation}
and consequently
\begin{equation}
  \varphi(\kappa,e,\omega;v_p) = \frac{(1-e^2)^{3/2}}{2 \pi \kappa}
  \frac{1}{\left| \sin(\phi+\omega) \right| ( 1 + e \cos \phi)^2}.
\end{equation}
The specific binary LOSVD is a function solely of $v_p$. To eliminate the
phase angle $\phi$ from this expression, equation (\ref{velo}) must be
inverted~:
\begin{equation}
\cos(\phi+\omega) = \frac{v_p}{\kappa} - e \cos \omega.
\end{equation}
Depending on the sign of $\sin(\phi+\omega)$, there are two solutions, $\varphi_+$ and $\varphi_-$~:
\begin{eqnarray}
  \varphi_\pm(\kappa,e,\omega;v_p) &=& \frac{1}{2 \pi \kappa} \frac{(1-e^2)^{3/2}}
  {\sqrt{1 - {\cal V}^2}} \times \nonumber \\ 
&& \hspace{-1em} \left( 1 + e {\cal V} \cos \omega \pm e
    \sqrt{1 - {\cal V}^2} \sin \omega \right)^{-2}
\end{eqnarray}
with ${\cal V} = v_p/\kappa - e \cos \omega$. This is a consequence
of the fact that a star on an elliptical orbit will obtain the
projected velocity $v_p$ twice during each revolution. The LOSVD is
then just the sum of $\varphi_+$ and $\varphi_-$~:
\begin{eqnarray}
  \varphi(\kappa,e,\omega;v_p) &=& \frac{1}{\pi \kappa} \frac{(1-e^2)^{3/2}}
  {\sqrt{1 - {\cal V}^2}} \times \nonumber \\
  && \hspace{-4em} \frac{1 + e^2 \sin^2 \omega + 2 e {\cal V} \cos \omega + e^2
    {\cal V}^2 \cos(2 \omega)}{\left( 1 - e^2 \sin^2 \omega + 2 e
      {\cal V} \cos \omega + e^2 {\cal V}^2 \right)^2}. \label{intrine}
\end{eqnarray}
In the case of a circular orbit -- i.e. $e=0$ -- this reduces to the
simple expression
\begin{eqnarray}
  \varphi(\kappa;v_p) &=& \frac{1}{\pi \kappa} \frac{1}{\sqrt{ 1 -
      \left( \frac{v_p}{\kappa} \right)^2}}, \label{intrine0}
\end{eqnarray}
the LOSVD of a harmonic oscillator. The specific binary LOSVD is
presented for various values of $\omega$ and $e$ in Figure
\ref{phi_intr}. The LOSVD is high around the extreme velocities where
$dv_p/dt=0$ (i.e.  around $v_p = \kappa(\pm 1 + e \cos \omega)$) and around
the apocenter velocity $v_p = (e-1) \kappa \cos \omega$, a direct
corollary of Kepler's second law.

\section{The velocity moments of the specific binary LOSVD}

The $n^{\rm th}$ velocity moment of the specific binary LOSVD is
defined as
\begin{eqnarray}
  \mu_{n} &=& \int v_p^n \,
  \varphi(v_p) \, dv_p  = \frac{1}{T} \int_0^T v_p^n \,dt\nonumber \\
  &=& \frac{\kappa^n}{2 \pi} (1-e^2)^{3/2} \int_0^{2 \pi} \frac{\left(
      \cos(\phi+\omega) + e \cos \omega \right)^{n}}{(1+e \cos
    \phi)^2} d\phi.
\end{eqnarray}
All powers in the integrand can be expanded, yielding
\begin{eqnarray}
  \mu_{n} &=& \frac{\kappa^n}{2 \pi} (1-e^2)^{3/2} \sum_{k=0}^{n} {n
    \choose k} (e \cos \omega)^{n-k} \times \nonumber \\
&& \hspace{-2em} \sum_{l=0}^k {k \choose l} (\cos \omega)^l
  (\sin \omega)^{k-l}
  (-1)^{k+l} \sum_{m=0}^\infty {-2 \choose m} e^m \times \nonumber \\
  && \int_0^{2 \pi} (\cos \phi)^{l+m} (\sin \phi)^{k-l}\,
  d\phi.
\end{eqnarray}
For the integral to be non-zero, one condition is that $k-l=2r$ with
$r$ an integer. The other condition is that $k+m-2r$ must be even.
Hence, $k$ an $m$ must both be either even or odd. This leads to the 
following expression for the $n^{\rm th}$ velocity moment~:
\begin{eqnarray}
  \mu_{n} &=& \frac{\kappa^n}{\pi} (1-e^2)^{3/2} \sum_{k=0}^{n} {n
    \choose k} \times \nonumber \\
  && \sum_{r=0}^{[k/2]} {k \choose k-2r} (\cos
  \omega)^{n-2r}  (\sin \omega)^{2r} \times \nonumber \\
  && \hspace{-1em} \sum_{m \ge 0}^{\hspace{.8cm}(k)} {-2 \choose m}
  e^{m+n-k} \frac{\Gamma(\frac{k+m+1}{2}-r)
    \Gamma(r+\frac{1}{2})}{\Gamma(\frac{k+m}{2}+1)}. 
\end{eqnarray}
Here, $[k/2]$ stands for the largest integer that is less than $k/2$
and $\sum_{m \ge 0}^{(k)}$ indicates that the summation runs only over
those $m$ that have the same parity as $k$. $\Gamma$ is Euler's Gamma
function (see e.g. Gradshteyn \& Ryzhik \cite{gry}, also for the other
special functions appearing in this paper). Interchanging the
summations, this expression can be rewritten in a more elegant form
\begin{eqnarray}
  \mu_{n} &=& \frac{(\kappa \cos \omega) ^n}{\sqrt{\pi}} (1-e^2)^{3/2}
  \sum_{k=0}^n {n \choose k} \!\!\!\!  \sum_{m \ge
    0}^{\hspace{.8cm}(k)} {-2 \choose m} \times \nonumber \\
  && e^{m+n-k} \frac{\Gamma\left(\frac{m+k+1}{2}\right)}{\Gamma\left(
      \frac{k+m}{2}+1 \right)} \times \nonumber \\
  && _2F_1\left( -\frac{k}{2},\frac{1-k}{2};\frac{1-k-m}{2};-
    \tan^2\omega \right). \label{mome}
\end{eqnarray}
For $e=0$, only the terms with $m+n-k=0$ survive. This condition can only 
be satisfied for even moments and if $m=0$ and $k=n$. Thus, if we
substitute $2n$ for $n$, and make use of the fact that
\begin{eqnarray}
  \cos^{2n} \omega \,\, _2F_1\left(-n,\frac{1}{2}-n;\frac{1}{2}-n;-
    \tan^2\omega \right) = 1, 
\end{eqnarray}
then (\ref{mome}) reduces to
\begin{eqnarray}
  \mu_{2n} &=& \frac{1}{\pi} \left( \frac{\mu_p}{a_p}\sin i^2 \right)^n
  B\left(n+\frac{1}{2},\frac{1}{2} \right) \label{mome0}
\end{eqnarray}
with $B$ Euler's beta function.

\section{The binary LOSVD for circular orbits}

The binary LOSVD $\tilde{\varphi}(M;v_p)$ gives the probability of
finding the primary star with a line-of-sight velocity in the interval
$[v_p-\Delta v_p/2,v_p+\Delta v_p/2]$. To obtain this quantity, the
specific binary LOSVD (\ref{intrine}) must be averaged over all
orbital parameters. As the calculations for the general case are quite
cumbersome, we only calculated the binary LOSVD in the approximation
that the members of all binaries revolve on circular orbits. This is
clearly not a very realistic assumption but it is instructive to have
at least this first approximation of the LOSVD. To obtain the binary
LOSVD, the specific binary LOSVD (\ref{intrine0}) must be averaged
over all possible values of the inclination $i$, the orbital radius
$a_p$ and the secondary mass $m$. To do this, we have to assume
distributions for each of these parameters.

\begin{figure}
\vspace{6.5cm}
\special{hscale=34 vscale=34 hsize=570 vsize=250 
         hoffset=-25 voffset=200 angle=-90 psfile="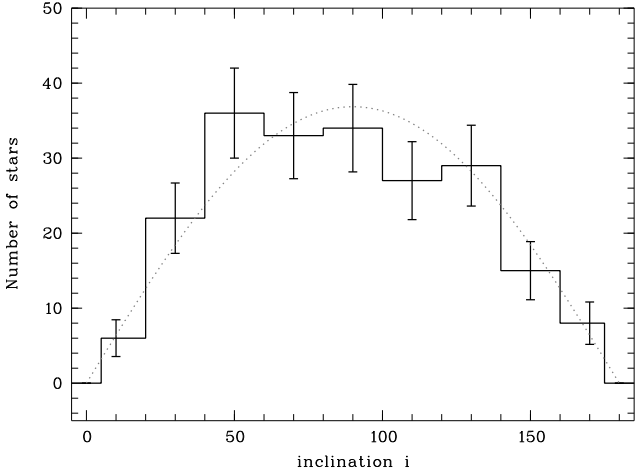"}
\caption{The distribution of the inclination $i$. The histogram
  shows the distribution extracted from Soederhjelm (1999) with
  the error-bars measuring the Poisson noise. Over-plotted (dotted line)
  is the expected distribution (\ref{disi}). \label{soei} }
\end{figure}
The orbital planes of all binaries can be expected to be distributed
randomly. Each direction of the normal of the orbital plane is then
equally plausible. Using the notations of Figure \ref{orbitps}, an
elementary solid angle around the orbital plane's normal is given by
$dS = \sin i \,di \,d\Omega$. Since the direction $S$ is uniformly
distributed
\begin{equation}
  \varphi(S)\,dS = \frac{1}{4 \pi} \,dS = \frac{1}{4 \pi} \sin i \, di
  \, d\Omega
\end{equation}
one obtains
\begin{equation}
  \varphi(i) = \int_0^{2 \pi} \varphi(S) \, d\Omega = \frac{1}{2} \sin
  i \hspace{1em} {\rm with}\hspace{1em}  i \in [0,\pi] \label{disi}
\end{equation}
as the distribution of the inclination. In Soederhjelm \cite{soe}, the
orbital parameters and masses of about 200 nearby visual binaries are
given.  The data are obtained from Hipparcos and ground-based
telescopes.  Figure \ref{soei} shows the distribution of the
inclination, extracted from this catalog, and clearly demonstrates the
validity of (\ref{disi}).

The distribution of the semi-major axis cannot readily be extracted
from a catalog. Catalogs of visual binaries usually only contain the
semi-major axis expressed in arcseconds since their distances are not
known. Catalogs of spectroscopic binaries do give the semi-major axis
expressed in e.g. AU but they tend to be strongly biased towards very
tight binaries since these will have the highest orbital velocities
and hence the largest and the easiest measurable Doppler-shifts.
However, it is natural to assume that more tightly bound binaries are
less easily disrupted by external influences. Hence, one would expect
to observe less binaries with lower binding energies. We therefore
adopt
\begin{eqnarray}
  \varphi(a)\,da &\propto& E^\gamma \nonumber \\
  &=& \frac{1-\gamma}{r_{\rm max}^{1-\gamma} - r_{\rm min}^{1-\gamma}}
  a^{-\gamma}\,da \label{disa}
\end{eqnarray}
as the distribution of $a$ (the orbital radius with respect to the
secondary). Here, $E$ is the binary's binding energy and $\gamma$ is a
real number. The closest gravitational few-body system at hand, our
planetary system, obeys (\ref{disa}) quite well with $\gamma \approx
1$.

The upper and lower bounds for $a$, $r_{\rm min}$ and
$r_{\rm max}$ respectively, can be estimated as follows. For a binary
to be stable, the mutual gravitational attraction between its members
should overcome the tidal forces exerted by the host galaxy. In other
words, if we denote the average mass of a star by $\overline{m}$~:
\begin{equation}
\frac{\overline{m}}{a^3} \ge \frac{2M(r)}{r^3}
\end{equation}
with $r$ the distance of the binary to the galaxy's center and $M(r)$
the total mass inside a sphere with radius $r$. This leads to the
following expression for the maximum orbital radius~:
\begin{equation}
r_{\rm max} \approx \sqrt[3]{\frac{3}{8 \pi \overline{n}}}
\end{equation}
with $\overline{n}$ the average number density of the stars inside the
radius $r$. For reasonable values of $\overline{n}$, $r_{\rm max}$ is
very large. For instance, $\overline{n}=1~{\rm pc}^{-3}$ leads to
$r_{\rm max} \approx 10^5~{\rm AU}$. Binaries can also be disrupted by
an encounter with a third star. For typical number densities, stars
approach each other to about 1500~AU once in a Hubble time as they
orbit in a galaxy's gravitational potential well. Of course, not every
such an encounter necessarily breaks up a binary pair. Hence, $r_{\rm
max}$ is at least a couple of thousand AU and can be as large as
100,000 AU. Our results turn about to be rather insensitive to the
value of $r_{\rm max}$. 

If the members of a binary are very close together, one of them can be
in the other's Roche lobe and be disrupted by tidal and centrifugal
forces. Back-of-the-envelope calculations based on typical masses and
radii of K-giants (with a mass around $1~{\rm M}_\odot$ and a radius
of approximately $10~{\rm R}_\odot$)--which one is most likely to
observe since they are so luminous -- give $r_{\rm min} \approx
15~{\rm R}_\odot$ as a fair mean value.

Due to (\ref{defa}), the distribution of $a_p$ is
\begin{eqnarray}
  \varphi(a_p)\,da_p &=& \frac{1-\gamma}{a_{\rm M}^{1-\gamma} -
    a_{\rm m}^{1-\gamma}} a_p^{-\gamma}\,da_p \label{disap}
\end{eqnarray}
with $a_{\rm m} = m/(m+M) r_{\rm min}$ and analogously for $a_{\rm
  M}$.

For the distribution of the secondary mass, we take the Salpeter
initial mass function
\begin{equation}
\varphi(m)\,dm = \frac{1-x}{m_2^{1-x}-m_1^{1-x}} m^{-x}\,dm \label{dism}
\end{equation}
with $x=2.35$. The lower bound for the mass distribution is taken to
be $m_1=0.08~{\rm M}_\odot$, the mass of the most light-weight stars
capable of nuclear fusion. The upper bound $m_2$ is treated as a free
parameter of the binary orbital distribution. Realistic values for
$m_2$ are around 1.25~M$_\odot$. This corresponds to stars with a
total life-time of 5-6~Gyr, comparable to the ages of the stellar
populations in Local Group dwarf galaxies (e.g. Smecker-Hane {\em et
al.} \cite{smh}). These distributions can be compared to those found
by Duquennoy~\&~Mayor \cite{dm} for a sample of 164 solar-type stars
in the solar neighborhood. For their sample, they find the following
distribution of the binary periods~:
\begin{equation}
\varphi(\log T)\, d \log T \propto \exp \left( - \frac{1}{2} \left(\frac{\log T -
\overline{\log T}}{\sigma_{\log T}} \right)^2 \right)\, d \log T
\end{equation}
with $\overline{\log T} = 4.8$ and $\sigma_{\log T}=2.3$ ($T$ is
expressed in days). Periods were found to lie in the range $-1 < \log
T < 10$. Using (\ref{disa}), (\ref{dism}) and $r_{\rm max} \approx
100,000$~AU (applicable if binaries are disrupted by tidal forces
exerted by the host galaxy), we find an approximately uniform
distribution of $\log T$ in the range $0 < \log T < 10$. Using $r_{\rm
max} \approx 2000$~AU (applicable if binaries are disrupted by a close
encounter with a third star), yields an approximately uniform
distribution of $\log T$ in the range $0 < \log T < 7.5$. As always,
the truth lies somewhere in between these two extremes. The Salpeter
law adopted here provides a reasonable fit to the mass-ratio
distribution although it over-estimates the number of low-mass
companions with $q<0.4$ found by Duquennoy~\&~Mayor.

We start with the integration of the specific binary LOSVD
(\ref{intrine0}) over the inclination, making use of the distribution
(\ref{disi}). The integration region must be limited to
\begin{equation}
  \arcsin t_0 < i < \pi -\arcsin t_0
\end{equation}
with $t_0 = \sqrt{ \frac{a_p v_p}{\mu_p} }$, to assure that the
argument of the square root in the denominator of (\ref{intrine0}) is
positive. This, of course, is because, for a given $a_p$ and $\mu_p$,
some binaries with a large inclination cannot reach the velocity $v_p$
and must be excluded from the integration. The binary LOSVD after this
integration becomes
\begin{eqnarray}
  \tilde{\varphi}(M;v_p) &=& \frac{1}{2 \pi v_p} \int_{\arcsin t_0}^{\pi -
    \arcsin
    t_0} \frac{\sin i}{\sqrt{ \frac{1}{t_0^2}\sin^2i -1 }}\, di \nonumber \\
  &=& \frac{1}{2} \sqrt{ \frac{a_p}{\mu_p} }. \label{l2}
\end{eqnarray}

We employ the distribution function (\ref{disap}) to average formula
(\ref{l2}) over the semi-major axis. The integration region is now
limited to
\begin{equation}
a_{\rm m} < a_p < \min \left( a_{\rm M},\frac{\mu_p}{v_p^2} \right) .
\end{equation}
This ensures the positivity of the binary LOSVD by excluding all
binaries that, for a given $\mu_p$, cannot reach the velocity $v_p$.
The binary LOSVD at this stage of the integrations takes the form
\begin{eqnarray}
  \tilde{\varphi}(M;v_p) &=& \frac{1-\gamma}{3-2\gamma}
  \frac{1}{r_{\rm max}^{1-\gamma} - r_{\rm min}^{1-\gamma}}
  \frac{1}{\sqrt{G}} m^{\gamma-5/2}
  (m+M)^{2-\gamma} \times \nonumber \\
  && \left[ \left( \min \left( a_{\rm M},\frac{\mu_p}{v_p^2} \right)
    \right)^{\frac{3}{2}-\gamma} - a_{\rm m}^{\frac{3}{2}-\gamma}
  \right].
\end{eqnarray}

The condition $a_{\rm M} \le \frac{\mu_p}{v_p^2}$ is satisfied if
\begin{equation}
  m \ge m_0 = \frac{M}{2} \left( \frac{r_{\rm max}v_p^2}{GM} + \sqrt{
      \frac{r_{\rm max}v_p^2}{GM} \left( 4 + \frac{r_{\rm
            max}v_p^2}{GM} \right)} \right).
\end{equation}
In that case, there is no danger of the integrand becoming negative.
If $m<m_0$ however, the integration region must be limited to
\begin{equation}
  m \ge m_3 = \frac{M}{2} \left( \frac{r_{\rm min}v_p^2}{GM} + \sqrt{
      \frac{r_{\rm min}v_p^2}{GM} \left( 4 + \frac{r_{\rm
            min}v_p^2}{GM} \right)} \right)
\end{equation}
if $\gamma < 3/2$. Clearly, $m_3 << m_0$ if $v_p > 0$. The integration
over the secondary mass involves three basic integrals. The first one
is
\begin{eqnarray}
  I_1(x_1,x_2) &=& \frac{1-\gamma}{3-2\gamma} \frac{r_{\rm
      max}^{3/2-\gamma}}{r_{\rm max}^{1-\gamma} - r_{\rm
      min}^{1-\gamma}} \frac{1}{\sqrt{G}}
  \frac{1-x}{m_2^{1-x}-m_1^{1-x}} \times \nonumber \\
&& \hspace{2em}  \int_{x_1M}^{x_2M} m^{-x-1} (m+M)^{1/2}\,dm \nonumber \\
  &=& \frac{1-\gamma}{3-2\gamma} \frac{r_{\rm
      max}^{3/2-\gamma}}{r_{\rm max}^{1-\gamma} - r_{\rm
      min}^{1-\gamma}} \frac{1}{\sqrt{G}}
  \frac{1-x}{m_2^{1-x}-m_1^{1-x}} \times \nonumber \\
&& \hspace{2em} M^{1/2-x} F\left(-x-1,\frac{1}{2};x_1,x_2\right).
\end{eqnarray}
with 
\begin{eqnarray}
  F(a,b;x_1,x_2) &=& \int_{x_1}^{x_2} x^a (1+x)^b\,dx, \,\,\,\,\,
  \mbox{\rm $x_1 < x_2$ }  \nonumber \\
    &=&  {\cal F}\left(a,b;x_1,x_2\right) \,\,\, \mbox{\rm if $x_2 \le 1$} \nonumber \\
    &=& {\cal F}\left(a,b;x_1,1\right)+ \nonumber \\
    && \hspace{1em} {\cal F}\left(-a-b-2,b;\frac{1}{x_2},1\right) \nonumber \\
    && \hspace{4em} \mbox{\rm if $x_2
      > 1$ but $x_1 \le 1$  } \nonumber \\
    &=& {\cal F}\left(-a-b-2,b;\frac{1}{x_2},\frac{1}{x_1} \right)
    \nonumber \\
    && \hspace{4em} \mbox{\rm if $x_2 > 1$ and $x_1 > 1$ }
\end{eqnarray}
and
\begin{eqnarray}
  {\cal F}\left(a,b;x_1,x_2\right) &=& \frac{x_2^{a+1}}{a+1}
  \,_2F_1(-b,1+a;2+a;-x_2) \nonumber \\
&&  - \frac{x_1^{a+1}}{a+1}
  \,_2F_1(-b,1+a;2+a;-x_1).
\end{eqnarray}
Here, $_2F_1$ is the hypergeometric function. The second integral is
completely analogous~:
\begin{eqnarray}
  I_2(x_1,x_2) &=& - \frac{1-\gamma}{3-2\gamma} \frac{r_{\rm
      min}^{3/2-\gamma}}{r_{\rm max}^{1-\gamma} - r_{\rm
      min}^{1-\gamma}} \frac{1}{\sqrt{G}}
  \frac{1-x}{m_2^{1-x}-m_1^{1-x}} \times \nonumber \\
&& \hspace{2em} 
  \int_{x_1M}^{x_2M} m^{-x-1} (m+M)^{1/2}\,dm \nonumber \\
  &=& - \frac{1-\gamma}{3-2\gamma} \frac{r_{\rm
      min}^{3/2-\gamma}}{r_{\rm max}^{1-\gamma} - r_{\rm
      min}^{1-\gamma}} \frac{1}{\sqrt{G}}
  \frac{1-x}{m_2^{1-x}-m_1^{1-x}}  \times \nonumber \\
&& \hspace{2em} M^{1/2-x}F\left(-x-1,\frac{1}{2};x_1,x_2\right).
\end{eqnarray}
The third basic integral is
\begin{eqnarray}
  I_3(v_p,x_1,x_2) &=& \frac{1-\gamma}{3-2\gamma} \frac{1}{r_{\rm
      max}^{1-\gamma} - r_{\rm min}^{1-\gamma}}
  \frac{1-x}{m_2^{1-x}-m_1^{1-x}}  \times \nonumber \\
&& \hspace{-6em}  \frac{1}{\sqrt{G}} \left( \frac{G}{v_p^2}
  \right)^{3/2-\gamma} \int_{x_1M}^{x_2M} m^{-2\gamma-x+2}
  (m+M)^{\gamma-1}\,  dm \nonumber \\
  &=& \frac{1-\gamma}{3-2\gamma} \frac{1}{r_{\rm
      max}^{1-\gamma} - r_{\rm min}^{1-\gamma}}
  \frac{1-x}{m_2^{1-x}-m_1^{1-x}}  \times \nonumber \\
&& \hspace{-8em}   \frac{M^{1/2-x}}{\sqrt{G}} \left( \frac{GM}{v_p^2}
  \right)^{3/2-\gamma} F\left(-2\gamma-x+2,\gamma-1;x_1,x_2\right).
\end{eqnarray}
In each of these integrals, the factor 
\begin{equation}
{\cal Q} = \frac{1-\gamma}{3-2\gamma} \frac{1}{r_{\rm
      max}^{1-\gamma} - r_{\rm min}^{1-\gamma}} \frac{M^{1/2-x}}{\sqrt{G}}
  \frac{1-x}{m_2^{1-x}-m_1^{1-x}}
\end{equation}
can be taken out. We are then left with the following three building
blocks~:
\begin{eqnarray}
{\cal I}_1(x_1,x_2) &=& r_{\rm max}^{3/2-\gamma}F\left(-x-1,\frac{1}{2};x_1,x_2\right), \nonumber \\
{\cal I}_2(x_1,x_2) &=& - r_{\rm min}^{3/2-\gamma}F\left(-x-1,\frac{1}{2};x_1,x_2\right), \nonumber \\
{\cal I}_3(v_p,x_1,x_2) &=& \left( \frac{GM}{v_p^2}
  \right)^{3/2-\gamma} \times \nonumber \\
&& \hspace{1em} F\left(-2\gamma-x+2,\gamma-1;x_1,x_2\right).
\end{eqnarray}
The binary LOSVD, averaged over all orbital parameters, finally takes
the following form~:
\begin{eqnarray}
  \tilde{\varphi}(M;v_p) &=& {\cal Q} \left[ {\cal I}_1\left(\max
      \left\{
        \frac{m_1}{M}, \frac{m_0}{M}\right\} ,\frac{m_2}{M} \right) \right.+ \nonumber \\
  && \hspace{-1em} \left. {\cal I}_2\left( \max \left\{ \frac{m_3}{M},
        \frac{m_1}{M}\right\},
      \frac{m_2}{M} \right) \right. + \nonumber \\
  && \hspace{-4em} \left.  {\cal I}_3\left(v_p, \max \left\{
        \frac{m_3}{M},\frac{m_1}{M} \right\}, \min \left\{
        \frac{m_2}{M},\frac{m_0}{M} \right\} \right) \right].
  \label{losvd}
\end{eqnarray}
\begin{figure}
\vspace{7.5cm}
\special{hscale=50 vscale=50 hsize=230 vsize=210 
         hoffset=-35 voffset=275 angle=-90 psfile="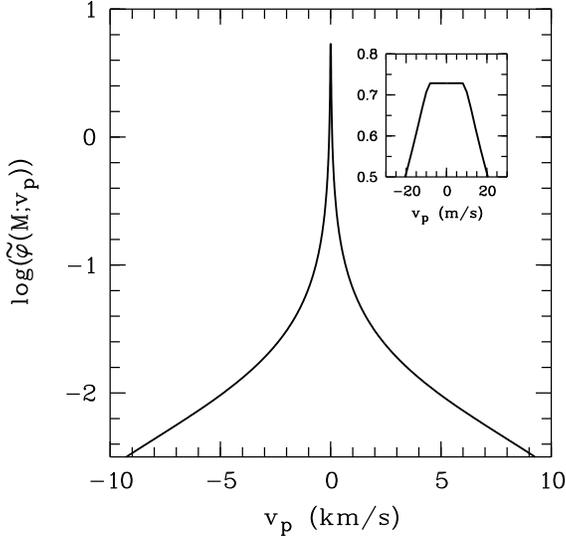"}
\caption{The logarithm of the binary LOSVD $\tilde{\varphi}(M;v_p)$ 
  for circular orbits. The primary mass is $M=0.8$~M$_\odot$, the
  secondary mass is averaged over the interval $0.08 - 1.5$~M$_\odot$,
  $r_{\rm min}=15~{\rm R}_\odot$, $r_{\rm max} = 101559.1~\,{\rm AU}
  \approx 0.5$~pc ($\overline{n}=1~{\rm pc}^{-3}$), $\gamma=1$ and
  $x=2.35$.  The inset shows the flat top of the LOSVD for very small
  velocities.  For the parameter values adopted here, the LOSVD is
  constant for velocities smaller than $v_1 \approx 7$~m/s.
  \label{losvdps} }
\end{figure}
As can be expected, the LOSVD goes to zero for $m_3 \ge m_2$ or
equivalently for
\begin{equation}
v_p \ge v_3 = \sqrt{ \frac{G m_2^2}{(m_2+M) r_{\rm min}}},
\end{equation}
the highest line-of-sight velocity any binary can reach. As long as
$m_0 < m_1$, or equivalently for
\begin{equation}
v_p \le v_1 = \sqrt{ \frac{G m_1^2}{(m_1+M) r_{\rm max}}},
\end{equation}
the LOSVD is velocity independent. This is the highest velocity that
the slowest binaries, i.e. the ones with the lowest secondary mass and
the largest orbital radius, can reach. Usually, $r_{\rm max}$ is very
large and consequently, $v_1$ will be a very small velocity, of the
order of a few m/s (see Figure \ref{losvdps}).

To be exact, the binary LOSVD (\ref{losvd}) still needs to be
multiplied with the probability that the primary star is actually
observed and then be averaged over the primary mass. It is this final
function that quantifies the extra broadening of the absorption lines
due to the binary population. However, the probability of observing a
star of a particular spectral type depends both on its intrinsic
brightness and on the number of such stars that are around. As the
brightest stars are also the most scarce ones while the faintest stars
are extremely abundant, this probability will be sharply peaked. For
old to intermediate stellar populations, this peak corresponds to
early K giants on the tip of the giant branch. We will therefore not do
this last integration and instead use an average value for the primary
mass. Most authors use $M=0.8\,{\rm M}_\odot$ and this is the value we
will adopt here.

\section{The moments of the binary LOSVD for non-circular orbits} \label{noncircmom}

The binary LOSVD for elliptic orbits is not analytically tractable,
but its velocity moments, which we denote by $\tilde{\mu}_n$, are. We
set out with the expression (\ref{mome}) for the moments of the
specific binary LOSVD and average it over the argument of the
periastron $\omega$, the inclination $i$, the secondary mass $m$, the
semi-major axis $a_p$ and the eccentricity $e$. For $i$ and $m$, one
can use the distributions (\ref{disi}) and (\ref{dism}). The
distribution (\ref{disap}) for the semi-major axis is also still
applicable but care has to be taken with the integration limits~: the
bounds for $a_p$ are set by the conditions
\begin{eqnarray}
a(1+e) \le r_{\rm max}, \nonumber \\
a(1-e) \ge r_{\rm min},
\end{eqnarray}
which makes the integration limits of $a_p$ dependent of $e$.

\begin{figure}
\vspace{6.5cm}
\special{hscale=34 vscale=34 hsize=570 vsize=250 
         hoffset=-25 voffset=200 angle=-90 psfile="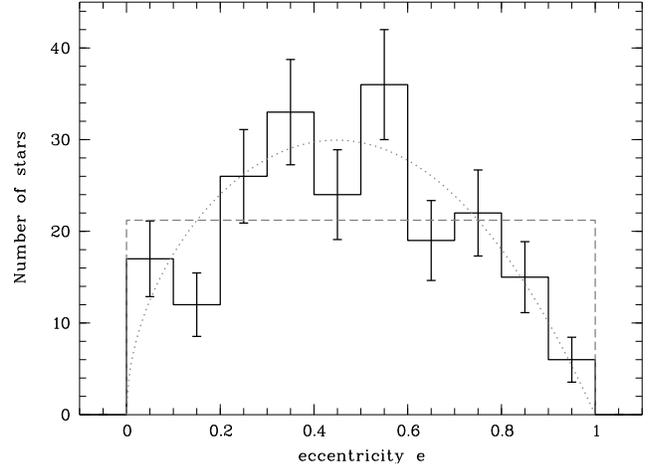"}
\caption{The distribution of the eccentricity $e$. The histogram
  shows the distribution extracted from Soederhjelm (1999) with
  the errorbars measuring the Poisson noise. Over-plotted in dotted
  lines is the distribution (\ref{dise}). The uniform distribution is
  plotted in dashed lines. \label{soee} }
\end{figure}
As can be seen in Figure \ref{soee}, the distribution of the
eccentricity of a sample of 200 visual binaries can be approximated fairly 
accurately by
\begin{equation}
  \varphi(e)\,de = \frac{(1+\alpha)e^\alpha (1-e^2)^\beta}{e_{\rm
      max}^{\alpha+1} \,_2F_1 \left(
      -\beta,\frac{1+\alpha}{2};\frac{3+\alpha}{2};e_{\rm max}^2
    \right)} \,de \label{dise}
\end{equation}
with $\alpha=0.5$, $\beta=1$ and maximum eccentricity $e_{\rm
  max}=1$. The eccentricity distribution found by Duquennoy~\&~Mayor
  \cite{dm} shows that tight binaries have more circular orbits than
  wide ones but, overall, their eccentricity distribution is similar
  to the one derived by Soederhjelm. In the remainder, we consider the
  parameters $\alpha$, $\beta$ and $e_{\rm max}$ to be fixed by the
  observations.  This distribution reaches a maximum for
\begin{equation}
e_{\rm top} = \sqrt{ \frac{\alpha}{2 \beta + \alpha}} \approx 0.45,
\end{equation}
for the parameter values adopted here. The uniform distribution
$\varphi(e) = 1/e_{\rm max}$, which is the distribution advocated by
Hargreaves {\em et al.} \cite{har} and Mateo {\em et al.}  \cite{mat},
is obtained by setting $\alpha = \beta = 0$.

The argument of the periastron $\omega$ is obviously distributed
randomly between $\omega=0$ and $\omega=2\pi$. Therefore,
\begin{equation}
\varphi(\omega)\,d\omega = \frac{1}{2\pi}\,d\omega.
\end{equation}

Averaging the $n^{\rm th}$ velocity moment of the specific binary LOSVD
(\ref{mome}) over $\omega$ involves the integral
\begin{equation}
  \frac{1}{2\pi} \int_0^{2\pi} (\cos \omega)^{n-2r} (\sin \omega)^{2r}
  \,d \omega
\end{equation}
which can only be non-zero if $n$ is even. We will therefore consider
only the even moments and make the transformation $n \rightarrow 2n$
in which case
\begin{eqnarray}
  \frac{1}{2\pi} \int_0^{2\pi} (\cos \omega)^{2n-2r} (\sin
  \omega)^{2r}
  \,d \omega &=& \nonumber \\
  && \hspace{-10em} \frac{1}{\pi}
  \frac{\Gamma(n-r+1/2)\Gamma(r+1/2)}{\Gamma(n+1)}.
\end{eqnarray}
The integrations over $i$ and $m$ are pretty straightforward and one
obtains the following expression for the moments $\tilde{\mu}_{2n}$~:
\begin{eqnarray}
  \tilde{\mu}_{2n} &=& \left( \frac{GM}{a} \right)^n
  \frac{1}{(2n+1)\sqrt{\pi}} \frac{1-x}{\left( \frac{m_2}{M}
    \right)^{1-x}-\left( \frac{m_1}{M} \right)^{1-x}} \times \nonumber \\
  && \hspace{-3em} F\left(2n-x,-n,\frac{m_1}{M};\frac{m_2}{M} \right)
  \sum_{k=0}^{2n} {2n \choose k} \!\!\!\!  \sum_{m \ge
    0}^{\hspace{.8cm}(k)} {-2 \choose m} \times \nonumber \\
&& \hspace{-3em} e^{m+2n-k} (1-e^2)^{3/2-n}
  \frac{\Gamma\left(\frac{m+k+1}{2}\right)}{\Gamma\left(
      \frac{k+m}{2}+1 \right)} \times \nonumber \\
&& \hspace{-3em} \,_3F_2 \left(
    -\frac{k}{2},\frac{1-k}{2},\frac{1}{2};\frac{1-k-m}{2},
    \frac{1}{2}-n;1 \right).
\end{eqnarray}

The next step is the integration over the semi-major axis $a$,
making use of (\ref{disa})~:
\begin{eqnarray}
  \frac{1-\gamma}{r_{\rm max}^{1-\gamma} - r_{\rm min}^{1-\gamma}}
  \int_{\frac{r_{\rm min}}{1-e}}^{\frac{r_{\rm max}}{1+e}}
  a^{-\gamma-n}\,da \!\! &=& \!\!\frac{1-\gamma}{1-\gamma-i} \frac{1}{r_{\rm
      max}^{1-\gamma} - r_{\rm min}^{1-\gamma}} \times \nonumber \\
&& \hspace{-10em} \left[ \left(
      \frac{r_{\rm max}}{1+e} \right)^{1-\gamma-n} - \left(
      \frac{r_{\rm min}}{1-e} \right)^{1-\gamma-n} \right].
\end{eqnarray}

Finally, we are left with the integration over all possible values of
the eccentricity. Making use of the lemma
\begin{eqnarray}
{\cal G}(a,b,c;x) && \nonumber \\
&&\hspace{-3em} = \int_0^x y^a (1+y)^b (1-y)^c \,dy \nonumber \\
&& \hspace{-3em} = \frac{x^{a+1}}{a+1} \sum_{m=0}^\infty
\frac{(-b)_m (a+1)_m}{(a+2)_m m!} (-x)^m \times
\nonumber \\
&& \,_2F_1(-c,a+m+1;a+m+2;x)
\end{eqnarray} 
one finds that~:
\begin{eqnarray}
  \tilde{\mu}_{2n} &=& \frac{(GM)^n}{\sqrt{\pi}}
  \frac{(1-\gamma)(1-x)}{(1-\gamma-n)(2n+1)} \frac{1}{r_{\rm
      max}^{1-\gamma} - r_{\rm min}^{1-\gamma}} \times \nonumber \\
  && \hspace{-4em} \frac{1}{\left( \frac{m_2}{M} \right)^{1-x}-\left(
      \frac{m_1}{M} \right)^{1-x}} \frac{1+\alpha}{e_{\rm
      max}^{\alpha+1} \,_2F_1 \left(
      -\beta,\frac{1+\alpha}{2};\frac{3+\alpha}{2};e_{\rm max}^2
    \right)} \times \nonumber \\
  && \hspace{-2em}  F\left(2n-x,-n;\frac{m_1}{M},\frac{m_2}{M} \right) \times \nonumber \\
  && \sum_{k=0}^{2n} {2n \choose k} \!\!\!\!  \sum_{m \ge
    0}^{\hspace{.8cm}(k)} {-2 \choose m}
  \frac{\Gamma\left(\frac{m+k+1}{2}\right)}{\Gamma\left(
      \frac{k+m}{2}+1 \right)} \times \nonumber \\
  &&\hspace{-3em} \,_3F_2 \left(
    -\frac{k}{2},\frac{1-k}{2},\frac{1}{2};\frac{1-k-m}{2},
    \frac{1}{2}-n;1 \right) \times \nonumber \\
  && \hspace{-4em}\left[ r_{\rm max}^{1-\gamma-n} {\cal G}
    \left(m+2n-k+\alpha,\frac{1}{2}+\gamma+\beta,\frac{3}{2}-n+\beta;e_{\rm
        max} \right) \right.  \nonumber \\ && \hspace{-4em}- \left.
    r_{\rm min}^{1-\gamma-n} {\cal G}
    \left(m+2n-k+\alpha,\frac{3}{2}-n+\beta, \right. \right. \nonumber \\
&& \hspace{8em} \left. \left. \frac{1}{2}+\gamma+\beta;e_{\rm
        max} \right) \right]. \label{momavere}
\end{eqnarray}
For $e_{\rm max}=0$, only the terms with $m+2n-k=0$ survive. This is
only possible if $m=0$ and $k=2n$ and (\ref{momavere}) reduces to
\begin{eqnarray}
\tilde{\mu}_{2n}   &=&
  \frac{(GM)^n}{2n+1}\frac{1-\gamma}{1-\gamma-n} \frac{r_{\rm
      max}^{1-\gamma-n} - r_{\rm min}^{1-\gamma-n}} {r_{\rm
      max}^{1-\gamma} - r_{\rm min}^{1-\gamma}} \times \nonumber \\
  && \hspace{-2em} \frac{1-x}{\left( \frac{m_2}{M}
    \right)^{1-x}-\left( \frac{m_1}{M} \right)^{1-x}}
  F\left(2n-x,-n;\frac{m_1}{M},\frac{m_2}{M} \right), \label{mombin0}
\end{eqnarray}
the expression for the moments of the LOSVD for
circular orbits. 

In the next paragraph, we discuss how the velocity dispersion of the
binary LOSVD for non-circular orbits depends on the distributions of
the different orbital parameters. Another important point that will be
addressed is the way the shape of the observed LOSVD depends on the
binary fraction.

\section{Discussion}

\subsection{The binary velocity dispersion as a function of the orbital 
  parameters}

\begin{figure}
\vspace{7cm}
\special{hscale=35 vscale=40 hsize=430 vsize=330 
         hoffset=-12 voffset=220 angle=-90 psfile="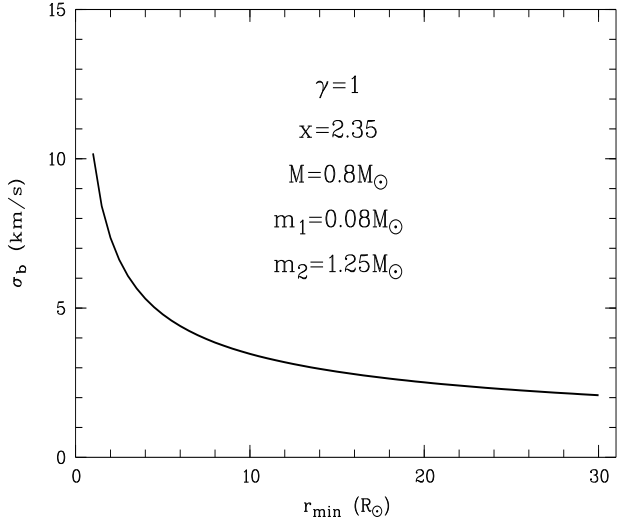"}
\caption{The binary velocity dispersion as a function of the inner 
  cutoff radius, $r_{\rm min}$. The values of the parameters are
  indicated in the plot.  \label{sigaminps} }
\end{figure}

\begin{figure}
\vspace{7cm}
\special{hscale=35 vscale=40 hsize=430 vsize=330 
         hoffset=-12 voffset=220 angle=-90 psfile="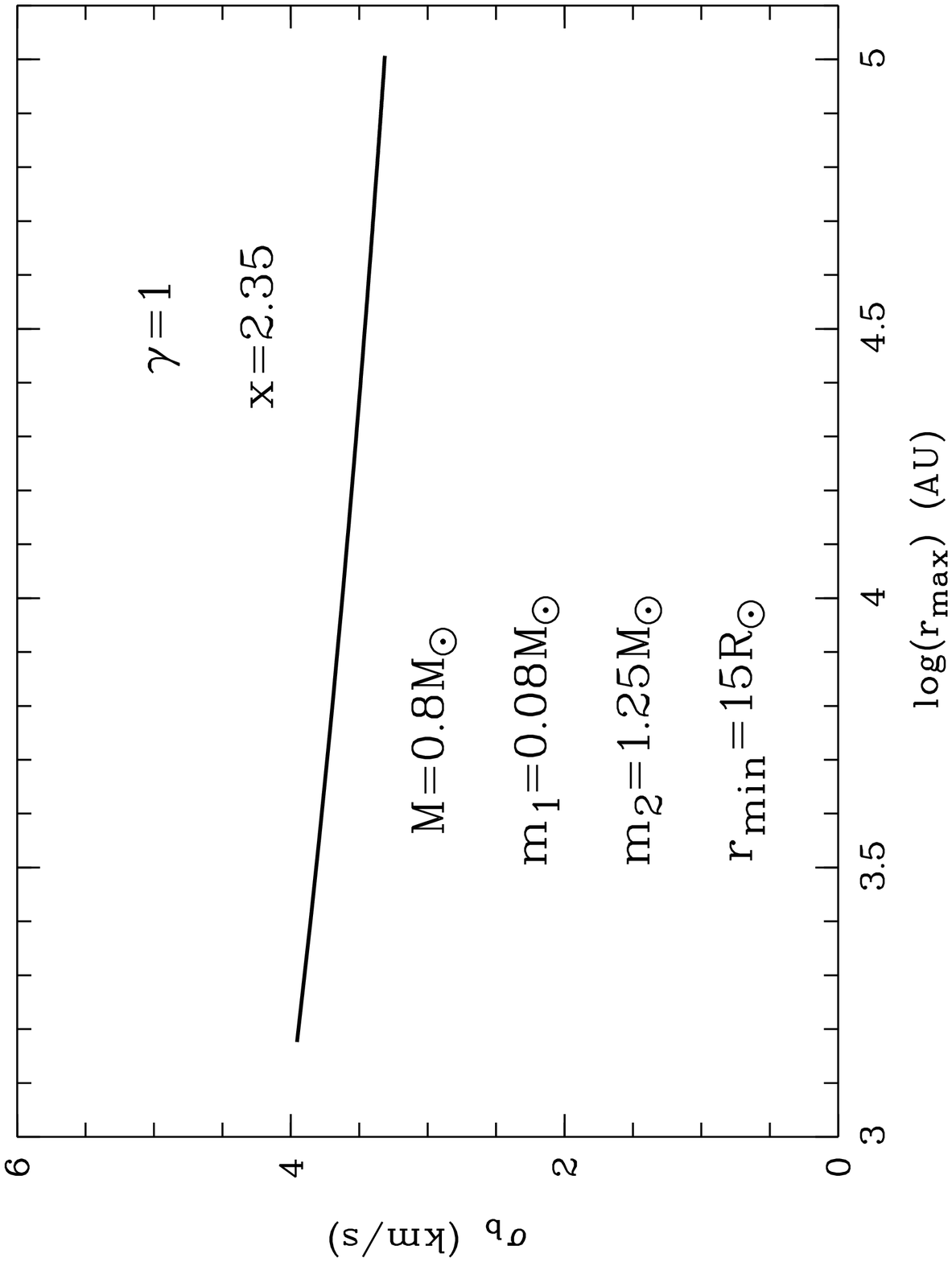"}
\caption{The binary velocity dispersion as a function of the outer
  cutoff radius, $r_{\rm max}$. The values of the parameters are
  indicated in the plot.  \label{sigamaxps} }
\end{figure}

\begin{figure}
\vspace{7cm}
\special{hscale=35 vscale=40 hsize=430 vsize=330 
         hoffset=-12 voffset=220 angle=-90 psfile="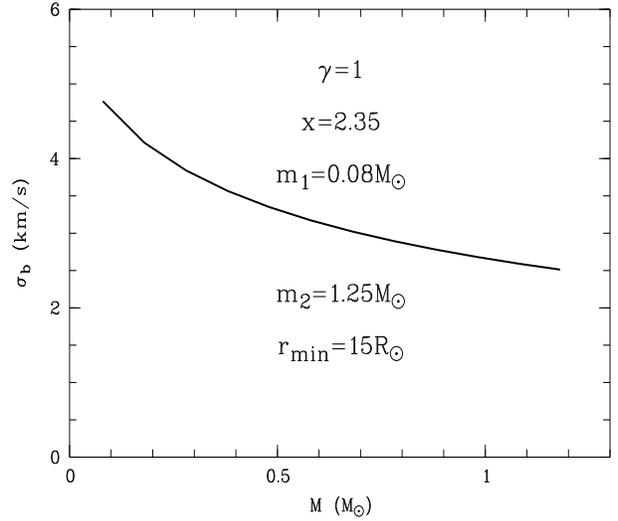"}
\caption{The binary velocity dispersion as a function of the primary mass $M$. 
\label{sigM}}
\end{figure}

\begin{figure}
\vspace{7cm}
\special{hscale=35 vscale=40 hsize=430 vsize=330 
         hoffset=-12 voffset=220 angle=-90 psfile="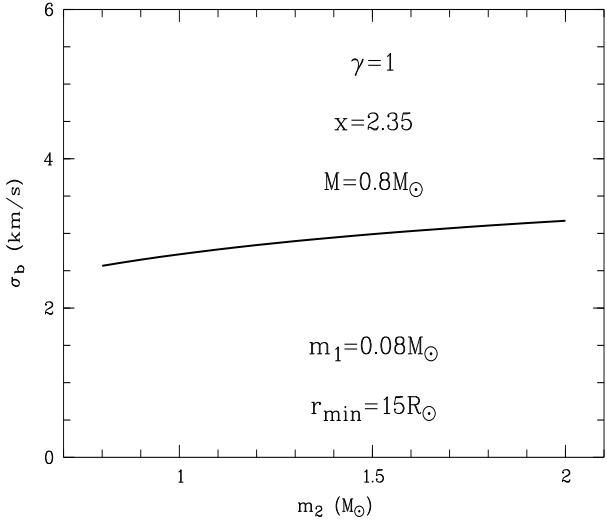"}
\caption{The binary velocity dispersion as a function of the upper mass cutoff 
  $m_2$. \label{sigm2}}
\end{figure}

\begin{figure}
\vspace{7cm}
\special{hscale=35 vscale=40 hsize=430 vsize=330 
         hoffset=-12 voffset=220 angle=-90 psfile="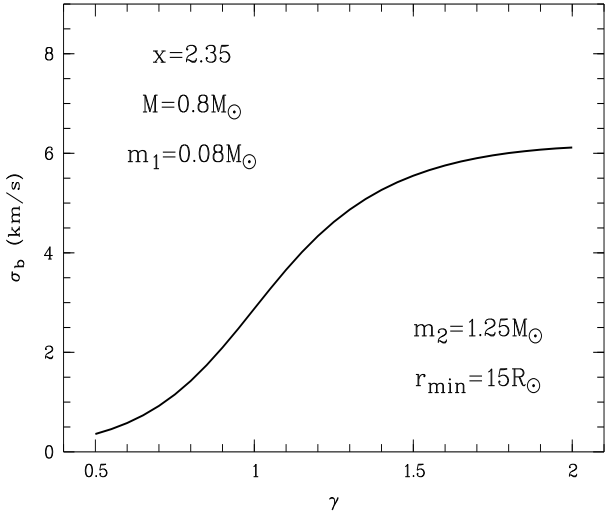"}
\caption{The binary velocity dispersion as a function of the exponent 
  $\gamma$ in the distribution of the major axis $a_p$.\label{siggam}
  }
\end{figure}

\begin{figure}
\vspace{7cm}
\special{hscale=35 vscale=40 hsize=430 vsize=330 
         hoffset=-12 voffset=220 angle=-90 psfile="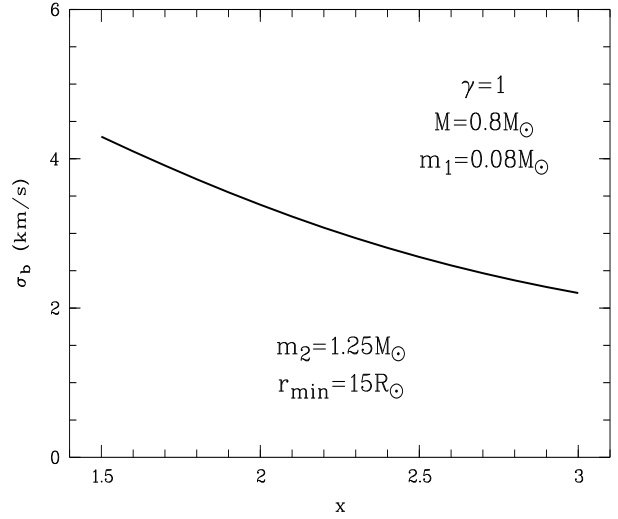"}
\caption{The binary velocity dispersion as a function of the exponent 
  $x$ in the Salpeter law for the secondary mass
  distribution.\label{sigx}}
\end{figure}

As is obvious from Figures \ref{sigaminps} to \ref{siggam}, the binary
velocity dispersion, i.e. the second order velocity moment
$\tilde{\mu}_2$, depends most sensitively on the inner cutoff radius
$r_{\rm min}$ and the exponent $\gamma$ in the distribution of the
semi-major axis $a_p$ and to a lesser degree on the primary mass $M$
and the exponent $x$ in the Salpeter IMF adopted as the distribution
of the secondary mass $m_2$. As the stars in a binary system are
allowed to approach each other more closely -- in other words, as
$r_{\rm min}$ is lowered -- their orbital velocities rise rapidly,
boosting the velocity dispersion of the binary population as a whole
to values as high as 10~km/s for $r_{\rm min}=1~{\rm R}_\odot$, as can
be seen in Figure \ref{sigaminps}. Since the bright giant stars that
one is most likely to observe have radii of the order of 10~${\rm
R}_\odot$, the velocity dispersion is limited to values below
4~km/s. The value of the outer cut-off radius, $r_{\rm max}$, does not
strongly affect the value of the binary dispersion. Higher $r_{\rm
max}$ values mean that more stars orbit on wide and consequently slow
orbits, causing the dispersion to be a bit lower. A higher value of
$\gamma$ causes stars to move on more tightly bound orbits, thus
boosting their velocities. This is clear from Figure \ref{siggam}. If
$x$, the exponent of the Salpeter IMF, is made larger, the secondary
masses will tend to be lower which also causes the primary's orbital
velocity and hence the velocity dispersion of the binary population as
a whole to be lower (see Figure \ref{sigx}). If $M$ is raised, the
center of mass will be closer to the primary, causing its orbital
velocity to drop slowly, as is observed in Figure \ref{sigM}. Larger
values for $m_2$ will produce heavier secondary stars. A slowly rising
velocity dispersion ensues, as can be seen in Figure \ref{sigm2}. The
same exercise was done with a uniform distribution for the
eccentricity. The outcome was essentially identical~: the binary
velocity dispersion differed at most by 0.1~km/s. Hence, our results
do not depend critically on the adopted distribution for the
eccentricity.

The values for the binary velocity dispersion obtained here are about
the same as those obtained by Hargreaves {\em et al.} although these
authors use Gaussian distributions for the ellipticity, secondary mass
and period, making it difficult to directly compare the results. Mateo
{\em et al.}  \cite{mat}, employing a Monte Carlo simulation with a
uniformly distributed ellipticity and secondary mass and a power law
distribution for the period, find higher velocity dispersions.
However, it is mentioned by Hargreaves {\em et al.} and by Olszewski
{\em et al.} that the values published by Mateo {\em et al.} are
overestimated due to a coding error. The corrected Mateo {\em et al.}
values however are in good agreement with the results obtained by
other authors and with those presented here.

In the following, we will adopt the model with
\begin{eqnarray}
\gamma &=& 1 \nonumber \\
x &=& 2.35  \nonumber \\
r_{\rm min} &=& 15~{\rm R}_\odot  \nonumber \\
m_1 &=& 0.08~{\rm M}_\odot  \nonumber \\
m_2 &=& 1.25~{\rm M}_\odot \nonumber \\
M &=& 0.8~{\rm M}_\odot
\end{eqnarray}
as our ``standard'' model. Its velocity dispersion is $\sigma_{\rm b}
= 2.87$~km/s and its kurtosis amounts to $\xi_{4, \rm{b}} = 86.89$. 


\subsection{The observed LOSVD}
\begin{figure}
\vspace{6cm}
\special{hscale=50 vscale=54 hsize=250 vsize=190 
         hoffset=-15 voffset=300 angle=-90 psfile="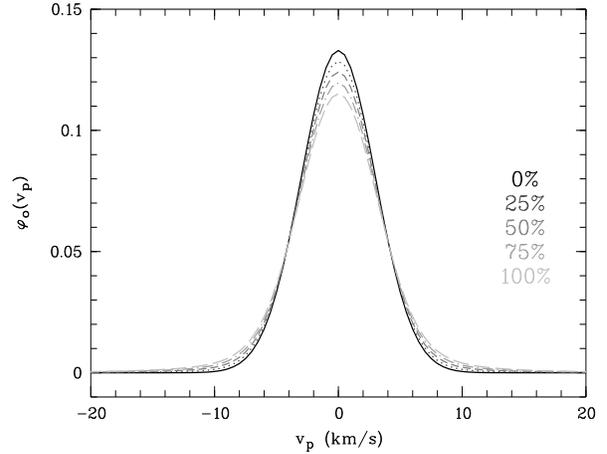"}
\caption{The observed LOSVD for different binary fractions 
  $\alpha$, indicated in the plot. The intrinsic LOSVD is a Gaussian
  with dispersion $\sigma_{\rm i} =3$~km/s, the binary LOSVD is given
  by (\ref{losvd}) with $M=0.8~{\rm M}_\odot$, $r_{\rm min}=15~{\rm
    R}_\odot$, $r_{\rm max} = 101559.1~\,{\rm AU}$
  ($\overline{n}=1~{\rm pc}^{-3}$), $\gamma=1$ and $x=2.35$.
  \label{loso}}
\end{figure}

The LOSVD that would actually be observed is the convolution of the
binary LOSVD (describing the motion of stars in binary systems) and
the intrinsic LOSVD of the galaxy (determined by the stars' orbital
motions through the galaxy). If we denote the intrinsic galaxy LOSVD
by $\varphi_{\rm i}(v_p)$ and the binary LOSVD by $\varphi_{\rm
  b}(v_p)$, then the observed LOSVD $\varphi_{\rm o}(v_p)$ can be
written as
\begin{eqnarray}
  \varphi_{\rm o}(v_p) &=& \int_{-\infty}^{+\infty} \varphi_{\rm
    i}(v_p-x)
  \left( (1-\alpha) \, \delta(x) + \alpha \, \varphi_{\rm b}(x) \right) \,dx \nonumber \\
  &=& (1-\alpha) \, \varphi_{\rm i}(v_p) + \alpha \int \varphi_{\rm
    i}(v_p-x) \, \varphi_{\rm b}(x) \,dx
\end{eqnarray} 
with $\delta(x)$ the Dirac delta function and $\alpha$ the binary
fraction, i.e. the fraction of visible (solar-mass) giant stars that
have a companion. As an indication, Duquennoy~\&~Mayor \cite{dm} find
$\alpha \approx 0.6$ for solar-mass stars in the solar neighborhood
with the same range of binary periods as adopted here. In Figure
\ref{loso}, the observed LOSVD is plotted for different binary
fractions. The intrinsic LOSVD is a Gaussian with dispersion
$\sigma_{\rm i}=3$km/s.  The binary LOSVD is given by (\ref{losvd})
with $M=0.8~{\rm M}_\odot$, $r_{\rm min}=15~{\rm R}_\odot$, $r_{\rm
max} = 101559.1~\,{\rm AU}$ ($\overline{n}=1~{\rm pc}^{-3}$),
$\gamma=1$ and $x=2.35$. The secondary mass is averaged over the
interval $0.08 - 1.5~{\rm M}_\odot$. The LOSVD becomes clearly more
peaked and develops broader high-velocity wings as the binary fraction
increases. A comprehensive way of assessing the change in the shape of
the observed LOSVD as a function of the binary fraction is to study
its moments. If the intrinsic galaxy LOSVD is a symmetric function of
the line-of-sight velocity, then one finds for the velocity dispersion
$\sigma_{\rm o}$ and the fourth order moment $\mu_{4,\rm o}$ of the
observed LOSVD~:
\begin{eqnarray}
  \sigma_{\rm o}^2 &=&  \sigma_{\rm i}^2 + \alpha \sigma_{\rm b}^2 \nonumber \\
  \mu_{4, \rm o} &=& \mu_{4,\rm i} + \alpha \mu_{4,\rm b} + 6 \alpha
  \sigma_{\rm i}^2 \sigma_{\rm b}^2
\end{eqnarray}
with $\sigma_{\rm i}^2$ and $\mu_{4,\rm i}$ respectively the second
and fourth order moments of the intrinsic galaxy LOSVD and
$\sigma_{\rm b}^2$ and $\mu_{4,\rm b}$ respectively the second and
fourth order moments of the binary LOSVD. The fact that we have the
fourth order moment of the binary LOSVD at our disposal allows us to
study not only the observed velocity dispersion but also the observed
LOSVD's kurtosis
\begin{equation}
\xi_{4,\rm o} = \frac{\mu_{4,\rm o}}{\sigma_{\rm o}^4}
\end{equation}
as a function of the binary fraction.
\begin{figure}
\vspace{7cm}
\special{hscale=35 vscale=40 hsize=430 vsize=330 
         hoffset=-12 voffset=220 angle=-90 psfile="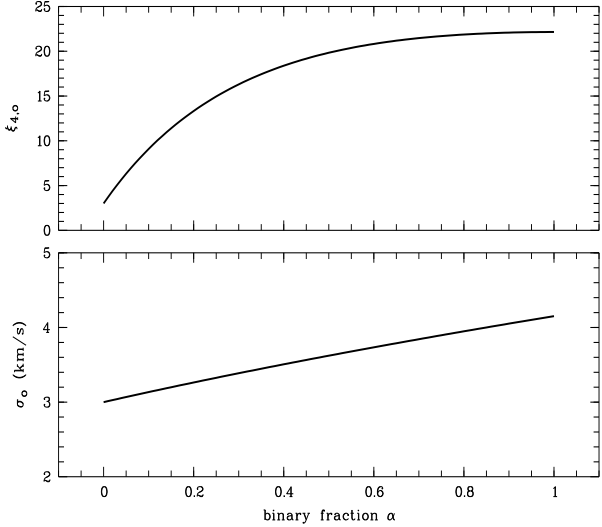"}
\caption{The velocity dispersion and kurtosis of
  the observed LOSVD as a function of the binary fraction $\alpha$.
  The intrinsic galaxy LOSVDs is a Gaussian with $\sigma_{\rm
    i}=3$~km/s.  The parameter values are those of the ``standard''
  model, i.e. with $\sigma_{\rm b} \approx 3$~km/s.
  \label{sigalfaps1}}
\end{figure}
\begin{figure}
\vspace{7cm}
\special{hscale=35 vscale=40 hsize=430 vsize=330 
         hoffset=-12 voffset=220 angle=-90 psfile="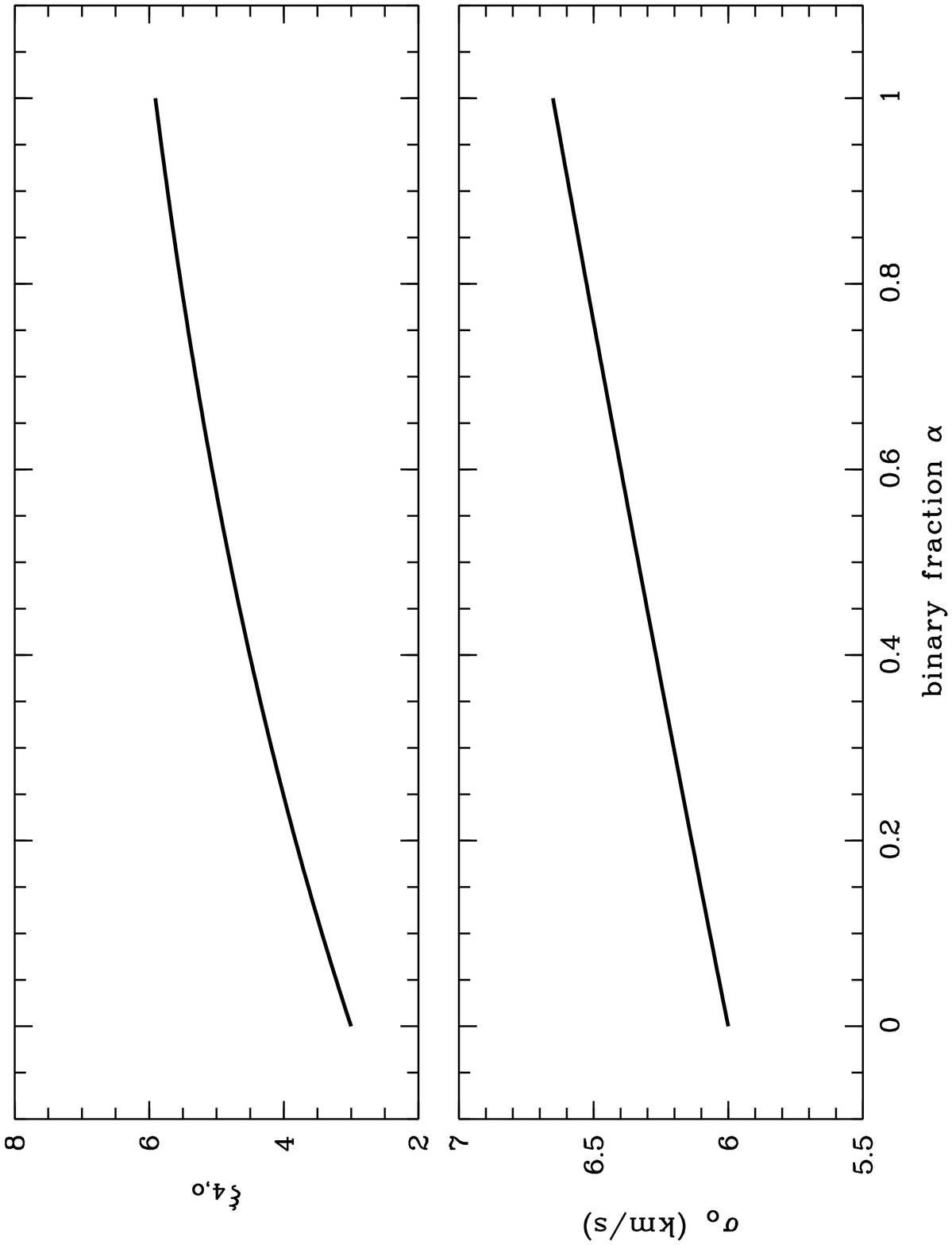"}
\caption{The velocity dispersion and kurtosis of
  the observed LOSVD as a function of the binary fraction $\alpha$.
  The intrinsic galaxy LOSVDs is a Gaussian with $\sigma_{\rm
    i}=6$~km/s. The parameter values are those of the ``standard''
  model, i.e. with $\sigma_{\rm b} \approx 3$~km/s.
  \label{sigalfaps2}}
\end{figure}
\begin{figure}
\vspace{7cm}
\special{hscale=35 vscale=40 hsize=430 vsize=330 
         hoffset=-12 voffset=220 angle=-90 psfile="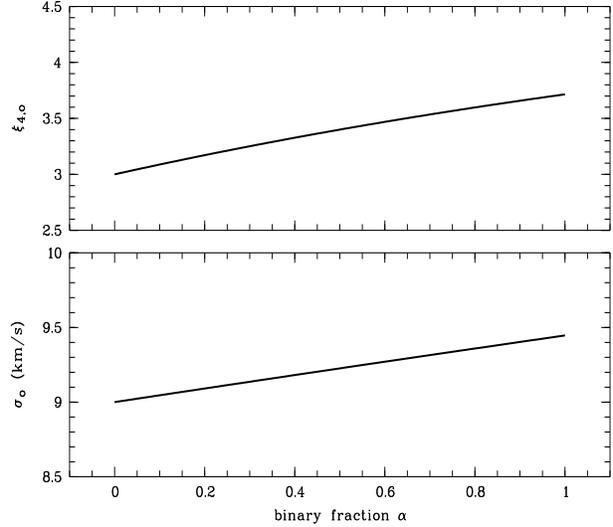"}
\caption{The velocity dispersion and kurtosis of
  the observed LOSVD as a function of the binary fraction $\alpha$.
  The intrinsic galaxy LOSVDs is a Gaussian with $\sigma_{\rm
    i}=9$~km/s. The parameter values are those of the ``standard''
  model, i.e. with $\sigma_{\rm b} \approx 3$~km/s.
  \label{sigalfaps3}}
\end{figure}

From Figures \ref{sigalfaps1}, \ref{sigalfaps2} and \ref{sigalfaps3},
it is clear that the velocity dispersion of only the most light-weight
galaxies will be appreciably affected by their binary population.
There, the velocity dispersion of the observed LOSVD is presented as a
function of the binary fraction for intrinsic Gaussian LOSVDs with
$\sigma_{\rm i}=3$~km/s, $\sigma_{\rm i}=6$~km/s and $\sigma_{\rm
  i}=9$~km/s. The kinematics of the binary population are those of the
``standard'' model. Only if the intrinsic velocity dispersion is of
the same order as the binary velocity dispersion, i.e. $\sigma_{\rm i}
\approx 3$~km/s, is there a noticeable change in the velocity
dispersion. A binary fraction of 60\% raises the velocity dispersion
of the $\sigma_{\rm i}=3$~km/s LOSVD with 24\%, that of the
$\sigma_{\rm i}=6$~km/s LOSVD with 7\% and that of the $\sigma_{\rm
  i}=9$~km/s LOSVD with 3\%. These Figures, however, yield a
surprising result~: whereas the influence of the binary population on
the observed velocity dispersion is limited to stellar systems with
intrinsic dispersions that are comparable to the binary velocity
dispersion, the shape of the observed LOSVD -- as quantified by the
kurtosis -- is appreciably affected for systems with intrinsic
dispersions as high as three times that of the binaries. The kurtosis
can be related to a quantity that is of more direct interest to
observers : the coefficient of the fourth order term in the expansion
of the LOSVD in Gauss-Hermite polynomials, $h_4$ (Gerhard \cite{ger},
van der Marel~\&~Franx \cite{vdm})~:
\begin{equation}
  h_4 \approx \frac{\xi_{4, \rm o}-3}{8 \sqrt{6}}.
\end{equation}
E.g. a binary fraction $\alpha=0.6$ raises the kurtosis of a
$\sigma_{\rm i}=6$~km/s Gaussian to 5.09 ($h_4 \approx 0.1$) and that
of a $\sigma_{\rm i}=9$~km/s Gaussian to 3.47 ($h_4 \approx 0.025$).
For stellar systems with lower intrinsic velocity dispersions, the
LOSVDs will be even more distinctly non-Gaussian.

\section{Conclusions}

The maximum additional velocity dispersion due to binary stars is
estimated at $\sigma_{\rm b} \approx 3$~km/s, in good agreement with
other authors. Only stellar systems with intrinsic velocity
dispersions comparable to this value will have observed velocity
dispersions that are noticeably affected by the presence of binaries,
i.e. dwarf Spheroidals such as those found in the Local Group
(Dekel~\&~Silk \cite{ds}), globular clusters, low-surface-brightness
disk galaxies (Bottema \cite{bot}) and the central regions of
nucleated dwarf ellipticals. Not only the velocity dispersion but also
the shape of the galaxy's LOSVD is altered. The presence of binaries
has a measurable effect on the kurtosis of the observed LOSVD for
stellar systems with a velocity distribution as high as 10~km/s.  For
stellar systems with lower velocity dispersions, the LOSVDs will be
even more distinctly non-Gaussian, being strongly peaked with broad
wings. This feature can mimic radial anisotropy which combined with
the enhanced velocity dispersion could lead to an over-estimation of
the dark matter content of dSphs and globulars.
  
If the kinematics of a galaxy are derived from the radial velocities
of discrete stars -- as is the case for the dwarf spheroidals in the
Local Group -- repeated observations can eliminate the binaries from
the star sample.  Methods that rely on integrated light spectra --
e.g. the crowded regions of globular clusters or dwarf galaxies
outside the Local Group -- will be plagued by the above mentioned
effects.

\end{document}